\documentclass[12pt,a4paper]{article}

\usepackage{amsmath}
\usepackage{amsfonts}
\usepackage{amssymb}
\usepackage{amsthm}

\newcommand{\dd}{{\mathrm d}}      

\newcommand{\R}{{\mathbb R}}   
\newcommand{\C}{{\mathbb C}}   
\newcommand{\HH}{{\mathbb H}}   

\DeclareMathOperator{\pf}{pf} 

\theoremstyle{definition}

\newtheorem{example}{Example}

\begin{document}

\title{Fermion integrals for finite spectral triples}

\author{John W. Barrett
\\ \\
School of Mathematical Sciences\\
University of Nottingham\\
University Park\\
Nottingham NG7 2RD, UK\\
\\
E-mail john.barrett@nottingham.ac.uk}

\date{2 August 2024}

\maketitle

\begin{abstract} Fermion functional integrals are calculated for the Dirac operator of a finite real spectral triple. Complex, real and chiral functional integrals are considered for each KO-dimension where they are non-trivial, and phase ambiguities in the definition are noted.
\end{abstract}

 \section{Introduction}
 Spectral triples provide a precise mathematical framework for the Dirac operator, allowing the generalisation to non-commutative spaces \cite{connesalainNoncommutativeGeometry1994}. In the traditional (commutative) case of a Dirac operator on a manifold, the formulation of the fermion functional integral is well known, and results in the computation of a regularised Pfaffian or determinant. 
 
 In this paper, the corresponding fermion integrals are calculated for finite spectral triples, that is, ones in which the vector space of spinor fields is finite-dimensional. Finite spectral triples are richer in the non-commutative setting and include the fuzzy sphere \cite{grosseDiracOperatorFuzzy1995} and the fuzzy torus \cite{barrettFiniteSpectralTriples2019}. Moreover, random finite spectral triples have been used to model random geometries \cite{barrettMonteCarloSimulations2016, glaserScalingBehaviourRandom2017, barrettSpectralEstimatorsFinite2019, azarfarRandomFiniteNoncommutative2024, khalkhaliPhaseTransitionRandom2020, hessamBootstrappingDiracEnsembles2022, khalkhaliSpectralStatisticsDirac2022, hessamDoubleScalingLimits2023}, reviewed in \cite{hessamNoncommutativeGeometryRandom2022}. These models have the potential to model quantum gravity and bosonic matter fields in a finite setting. Thus it is interesting to add fermion fields and the fermion functional integral to these models. 
 
 Three different types of functional integral are considered: 
 \begin{itemize}
 \item The complex integral, over a set of fermionic variables and their conjugates. 
 \item The real integral, over a set of fermionic variables (without conjugates).
 \item The chiral integral, a real integral where the spinors are restricted to a definite chirality.
 \end{itemize}
There are eight different types of spectral triple distinguished by the KO-dimension $s$, which is an integer mod 8. The results of this paper are to determine the non-trivial cases, giving a formula for fermion integral in terms of a Pfaffian or a determinant. In some cases there is a sign or a phase ambiguity. Finally, there are some remarks about the physical significance of the results.

\section{Finite real spectral triples}

A finite spectral triple has a finite-dimensional Hilbert space $\mathcal H$, on which acts a self-adjoint operator $D$, the Dirac operator, and an algebra $\mathcal A$. The eight different types of spectral triple are determined by the properties of a chirality operator $\Gamma$ on $\mathcal H$, which is Hermitian and squares to $1$, and an antilinear unitary operator $J$ on $\mathcal H$ called the real structure.

 For even parameter $s$, $D$ anticommutes with $\Gamma$. The elements of the $+1$ eigenspace of $\Gamma$ are called the left-handed fermions and the $-1$ eigenspace the right-handed fermions. For odd $s$, $\Gamma$ commutes with $D$ and one can decompose the spectral triple so that $\Gamma=1$ (or $-1$); the chirality operator is therefore not important.

The properties of $J$ are $J^2=\epsilon=\pm1$, $DJ=\epsilon'JD=\pm JD$ and $J\Gamma J^{-1}\Gamma=\epsilon''=\pm1$.  These three signs depend only on the value of $s$, as shown in Figure \ref{signtable} \cite{connesNoncommutativeGeometryReality1995}.
 \begin{figure}
$$\begin{tabular}{|l|rrrrrrrr|}
  \hline
  $s$ &0&1&2&3&4&5&6&7\\
\hline
$\epsilon$&1&1&$-1$&$-1$&$-1$&$-1$&1&1\\
$\epsilon'$&1&$-1$&1&1&1&$-1$&1&1\\
$\epsilon''$&1&1&$-1$&1&1&1&$-1$&1\\
  \hline
&$\R$&$\C$&$\HH$&$\HH$&$\HH$&$\C$&$\R$&$\R$\\
\hline
\end{tabular}$$
\caption{Table of signs for real spectral triples.}\label{signtable}
\end{figure}
The properties of $J$ mirror those for a real structure in a Clifford module and therefore arise also where the fermions are spinors of signature $(p,q)$, with $s=q-p \mod 8$ \cite{barrettMatrixGeometriesFuzzy2015}.  

There are some further axioms that relate the algebra to the Dirac operator, chirality operator and the real structure. However, these axioms are not needed in this paper, except to note that the Dirac operator is constrained by the algebra action. There are no such constraints if one takes $\mathcal A=\C$. Note that since the Hilbert space has a positive definite inner product, any symmetries are unitary and the spectral triples are therefore an abstraction of Euclidean signature physics.

\section{Complex fermion integral}
The simplest fermionic action is
\begin{equation}\langle\psi,D\psi\rangle\label {eq:fermionaction}
\end{equation}
using the Hermitian inner product from $\mathcal H$. The fermion integral is 
\begin{equation}Z=\int_{\psi\in\mathcal H}\int_{\overline \psi\in\mathcal H}e^{i\langle\psi,D\psi\rangle}\;\dd \overline\psi\,\dd \psi=\det iD.
\end{equation}
which is the standard result, though it is worth writing the detail for later use. It is called a complex integral because the field and its conjugate are integrated independently.

The integral is defined by picking an orthonormal basis $e_j$ of $\mathcal H$, writing $\psi=\sum_j\psi_j e_j$. The $\psi_j$ are the component functions with respect to the basis and are assumed to be anticommuting variables. Then evaluating 
\begin{equation}Z=\int e^{i\sum_{jk}\langle e_j,D e_k\rangle\overline\psi_j\psi_k}\;\dd \overline\psi_1\,\dd \psi_1\,\dd \overline\psi_2\,\dd \psi_2\ldots=\det [\langle e_j,iD e_k\rangle].
\end{equation}
This gives the determinant of the matrix on the right-hand side, which is the determinant of the operator $iD$. In particular, it is independent of the orthonormal basis chosen. 

If the spectrum is symmetric about zero, as it is for $s=0,1,2,4,5\text{ or }6$, then this determinant is the product of terms $(i\lambda)(-i\lambda)$, and possibly zeroes, and so is a non-negative real number. 

\section{Real fermion integral}\label{sec:real}

Define a new inner product on $\mathcal H$
\begin{equation} [\psi,\phi]=\langle J\psi,\phi\rangle.
\end{equation}
This is bilinear, and a calculation with commuting $\psi$ and $\phi$ shows that $[\psi,\phi]=\langle J\psi,\phi\rangle
=\langle J\phi,J^2\psi\rangle=\epsilon [\phi,\psi]$.
Including the Dirac operator gives 
$ [\psi,D\phi]=\epsilon[D\phi,\psi]=\epsilon' \epsilon [\phi,D\psi] $.

An action is defined for the anticommuting field $\psi$ as
\begin{equation}\frac12 [\psi,D\psi]=\frac12\langle J\psi,D\psi\rangle \label{eq:realaction}
\end{equation}
which is only non-zero as an element of the Grassman algebra if $\epsilon' \epsilon=-1$, due to the anticommutation of fermions. 

Since the action depends on $\psi$ but not its complex conjugate, the fermion integration is symbolically
\begin{equation}F=\int_{\psi\in\mathcal H}e^{\frac i2[\psi,D\psi]}\;\dd \psi.\label{realfermionintegral}
\end{equation}
It is called a real integral because the conjugate variables are absent (like a contour integral in complex analysis).
Picking an ordered orthonormal basis $e_j$, the precise definition of this integral is
\begin{equation}F=\int_{\psi\in\mathcal H}e^{\frac i2\sum_{jk}[e_j,De_k]\psi_j\psi_k}\;\dd \psi_1\,\dd\psi_2\ldots=\pf (iM)
\end{equation}
with $M$ the matrix with components $M_{jk}=[e_j,De_k]$. This is an antisymmetric matrix if $\epsilon\epsilon'=-1$,  
and so the KO-dimension can only be $s=1,2,3,4$ if the fermion integral is non-zero.

The Pfaffian of $M$ depends on the choice of basis: in a new basis determined by unitary matrix $U$ this is  
\begin{equation}\pf(UMU^T)=\det U\pf M.\label{changeofbasis}
\end{equation} 
It follows that $\vert\pf M\vert$ is independent of the basis. 

However the phase of the Pfaffian is ambiguous with a general choice of basis. To reduce this ambiguity, the strategy is to use the operators $J$ and $\Gamma$ to restrict the bases considered for the definition of the integral. In this way it is possible to reduce or even completely remove the ambiguity, depending on the case considered.

The complex conjugate of the matrix is determined by
\begin{equation}\overline{M_{jk}}= \overline{\langle Je_j,De_k\rangle}=\langle J^2e_j,JDe_k\rangle=\epsilon'[ Je_j,DJe_k].\label{eq:conjm}
\end{equation}
Further analysis depends on the sign $J^2=\epsilon$.

\paragraph{Case $\epsilon=1$.}
The only non-trivial case this applies to is $s=1$. The basis is chosen to be a set of real vectors $Je_k=e_k$.  Since $\epsilon'=-1$, \eqref{eq:conjm} shows that $M_{jk}$ is imaginary, and so $\pf iM$ is a real number. A change of basis matrix $U$ is real, and so $\det U=\pm1$. Thus the sign of the Pfaffian depends on the ordering of the basis, up to even permutation. Another way of seeing that there is only a sign ambiguity is that $(\pf M)^2=\det [\langle e_j,D e_k\rangle]=\det D$, which is independent of the basis.

\paragraph{Case $\epsilon=-1$.}
 In this quaternionic case $J^2=-1$, nontrivial for $s=2,3,4$, an orthonormal basis is chosen such that $e_{2k}=Je_{2k-1}$ for $k=1,\ldots,n/2$. Then $\epsilon'=1$ and $\overline M=VMV^T$ with $V$ the direct sum of the $2\times2$ blocks
\begin{equation}\begin{pmatrix}0&1\\-1&0\end{pmatrix},
\end{equation}
which has determinant $1$.
Thus $\pf M$ is real (and $\pf iM$ either imaginary or real depending on whether $n/2$ is odd or even). If also $e'_{2k}=Je'_{2k-1}$, the change of basis matrix $U$ preserves the antisymmetric form $[e_j,e_k]$, and so $\det U=1$. Thus $\pf M$ is independent of the choice of basis. This is the same as the conclusion reached in \cite[\S 3.4]{stoneGammaMatricesMajorana2022}.

The Pfaffian can be calculated using an orthonormal basis of eigenvectors of $D$. Let $e_1$ be any eigenvector of $D$, with eigenvalue $\lambda$. Then $e_2=Je_1$ is a linearly independent eigenvector with the same eigenvalue and so $M_{12}=-M_{21}=\lambda$, which is the contribution to the Pfaffian. Repeating this with further eigenvectors shows that $(\pf M)^2=\det(D)$. 

In the cases $s=2$ or $4$, the chirality operator $\Gamma$ determines further eigenvectors $e_3=\Gamma e_1$ and $e_4=Je_3$, both with eigenvalue $-\lambda$. Thus (assuming all eigenvalues are non-zero) $n/2$ is even and $\pf iM$ is a product of terms $(i\lambda)(-i\lambda)$, and is therefore positive. In fact, $\pf iM=\sqrt{\det D}$.

If $s=3$, the chirality operator is trivial. A $2\times 2$ example shows that $\pf M$ can have either sign.

\section{Chiral fermions}\label{sec:Real}
For $s$ even, a chiral model is achieved by splitting $\mathcal H=\mathcal H_+\oplus\mathcal H_-$ according to the eigenvalues of the chirality operator $\Gamma$ and using only vectors in $\mathcal H_+$, or only vectors in $\mathcal H_-$.

For action \eqref{eq:fermionaction} this does not work because $D$ maps $\mathcal H_+$ to $\mathcal H_-$. Putting $P_\pm=\frac12(\Gamma\pm1)$, this problem is explicitly that $\langle P_+\psi,DP_+\psi\rangle=0$. 

The chirality condition can be imposed in the action \eqref{eq:realaction} for the real integral
 if the KO-dimension $s=2$. This is because $[P_+\psi,D P_+\psi]$ is non-zero in this case, due to the fact that $\epsilon''=-1$. (The only other case satisfying this condition is $s=6$, but the action for this case is zero.)

The chiral fermionic integrals are
\begin{equation}F_+=\int_{\psi\in\mathcal H_+}e^{\frac i2[\psi,D\psi]}\;\dd \psi=\pf i M_+
\end{equation}
and
\begin{equation}F_-=\int_{\psi\in\mathcal H_-}e^{\frac i2[\psi,D\psi]}\;\dd \psi=\pf i M_-
\end{equation}
with the matrix $M_\pm$ the restriction of $M$ to a basis of $\mathcal H_\pm$.
Once again, this depends on the choice of basis and the ambiguity is a phase factor $\det U_+$ or $\det U_-$ determined by the analogue of \eqref{changeofbasis}.

\begin{example}The phase ambiguity can be eliminated by considering an integral that has two chiral fields. The (non-chiral) fermion integral \eqref{realfermionintegral} for $s=2$ can be considered a product of two chiral integrals
\begin{equation}F=F_+F_-\label{eq:chiralsplit}
\end{equation}
using a basis of vectors each having definite chirality. The corresponding phase factors $\det U_+$ and $\det U_-$ cancel for the left- and right-handed bases because they are related with $J$. Explicitly,
\begin{equation}\psi=\begin{pmatrix}\psi_L\\\psi_R\end{pmatrix}, \quad\Gamma\psi=\begin{pmatrix}\psi_L\\-\psi_R\end{pmatrix}, \quad J\psi=\begin{pmatrix}J_R\psi_R\\ J_L\psi_L\end{pmatrix},\quad D\psi=\begin{pmatrix}D_R\psi_R\\ D_L\psi_L\end{pmatrix}
\end{equation}
and the action is 
\begin{equation}
\frac12 [\psi,D\psi]=\frac12 [\psi_L,D_L\psi_L]+\frac12 [\psi_R,D_R\psi_R]
\end{equation}
giving the splitting \eqref{eq:chiralsplit} on integration.\end{example}

\begin{example} \label{ex2} A different example is obtained by starting with a real spectral triple $(\mathcal A,\mathcal H,D, J,\Gamma)$ for $s=4$.
The Hilbert space is doubled, setting $\mathcal H'=\mathcal H\oplus\mathcal H$ and
\begin{equation} \Gamma'=\begin{pmatrix}\Gamma&0\\0&-\Gamma\end{pmatrix},\quad\quad
D'=\begin{pmatrix}D&\overline\mu\Gamma\\  \mu\,\Gamma&D\end{pmatrix},\quad\quad
J'=\begin{pmatrix}0&J\\J&0\end{pmatrix}
\end{equation}
with a constant $\mu\in \C$.
One can check this gives a spectral triple $(\mathcal A,\mathcal H',D', J',\Gamma')$ of KO-dimension $s=2$. The chiral condition for this spectral triple is the restriction to $\psi=(\chi_L,\phi_R)\in\mathcal H'_+\cong\mathcal H_+\oplus\mathcal H_-$, giving the chiral action
\begin{equation}\frac12[\psi,D'\psi]'=[\phi_R,D\chi_L]-\frac{\overline\mu}2[\phi_R,\phi_R]+\frac{\mu}2[\chi_L,\chi_L]\label{eq:SM}
\end{equation}
where the right-hand side is the expansion into two chiral fermions in the original $s=4$ Hilbert space $\mathcal H$. The chiral fermion integral is defined without a phase ambiguity by using the bases of  both $\mathcal H_+$ and $\mathcal H_-$ that are determined by $J$, as in Case $\epsilon=-1$ of Section \ref{sec:Real}. These are well defined because in KO-dimension 4, $J$ maps $\mathcal H_+$ to itself and also $\mathcal H_-$  to itself.

The integral can be evaluated, giving
\begin{equation} F_+=\pf iM=\sqrt[4]{\det(D^2+|\mu|^2)}\, e^{i\theta I_D/2}\label{Fchiral}
\end{equation}
where $e^{i\theta}=i\mu/|\mu|$, and $I_D$ is the index of the Dirac operator \cite{barrettMatrixGeometriesFuzzy2015}. This is proved as follows. Let $\psi\in\mathcal H$ be a normalised eigenvector, $D\psi=\lambda \psi$. If $\lambda\ne0$, define 
 \begin{equation}
 \psi_1=\frac1{\sqrt2}(1+\Gamma)\psi,\quad \psi_2=J\psi_1,\quad \psi_3=\frac1{\sqrt2}(1-\Gamma)\psi,\quad \psi_4=J\psi_3.
 \end{equation}
 Then $e_1, e_2\in\mathcal H_+$ and $e_3, e_4\in\mathcal H_-$, and the part of the matrix $M$ for the basis vectors $e_1=(\psi_1,0)$, $e_2=(\psi_2,0)$, $e_3=(0,\psi_3)$, $e_4=(0,\psi_4)\in\mathcal H'_+$ is
 \begin{equation}M_\psi= \begin{pmatrix} 0&\mu&0&\lambda\\
 -\mu&0&-\lambda&0\\
 0&\lambda&0&-\overline\mu\\
 -\lambda&0&\overline\mu&0\end{pmatrix}, \quad \pf M_\psi=-\lambda^2-|\mu|^2
 \end{equation}
 If $\lambda=0$ and $\psi\in\mathcal H_+$ then putting $e_1=(\psi,0)$, $e_2=(J\psi,0)$, the matrix is
 \begin{equation}M_\psi= \begin{pmatrix} 
 0&\mu\\
 -\mu&0\\\end{pmatrix}, \quad \pf M_\psi=\mu
 \end{equation}
 and for $\psi\in\mathcal H_-$, $e_1=(0,\psi)$, $e_2=(0,J\psi)$, the matrix is
 \begin{equation}M_\psi= \begin{pmatrix} 
 0&-\overline\mu\\
 \overline\mu&0\\\end{pmatrix}, \quad \pf M_\psi=-\overline\mu
 \end{equation}
The whole matrix $M$ is a direct sum of these types of block, and taking the product of their Pfaffians gives \eqref{Fchiral}.
\end{example}

 \section{Discussion}
 Non-trivial fermion integrals exist for all $s$ in the complex case without any ambiguity, but in the case of the real fermion functional integral, only for $s=1$  with a sign ambiguity, or $s=2, 3$ or $4$ with no ambiguity. In the chiral case, only $s=2$ is possible, and there is in general a phase ambiguity. There are examples of the chiral construction where the ambiguity disappears if the Hilbert space is constructed from two chiral fermions. 
 
In a sense, the ambiguities are not fatal because one can just pick a basis and obtain a definite value for the integral, despite the fact that the problem is not uniquely specified in an abstract mathematical sense. However, the issue may resurface when considering ensembles, because all of the possible signs or phases are liable to be equally likely and give cancelling contributions to a partition function when summing over Dirac operators. 

Thus the models with no ambiguity are the most important ones. Although the spectral triples considered here are all finite, the results that can be formulated in terms of determinants lead to a definition in the infinite-dimensional case, using regularised determinants. The regularisation can even be with finite-dimensional approximations \cite{andrianovRegularizedFunctionalIntegral1982}.

The physical interpretation of the KO dimension is subtle. In the simplest cases, $s$ is determined by a Clifford module and so could be the dimension of an analogous differentiable manifold. However, this is far from always the case, for example, the simplest spectral triple for the fuzzy 2-sphere \cite {grosseDiracOperatorFuzzy1995} has $s=3$. In fact, in general one can tensor the spectral triple for a manifold of dimension $d$ with an `internal space' of KO dimension $s$ to get an overall spectral triple with KO dimension $s'=d+s \mod 8$. This construction was used for the standard model of particle physics \cite{connesNoncommutativeGeometryStandard2006} with an internal space of KO dimension 6 and $d=4$ to give the chiral theory with $s'=2$. 

This construction is simplified and abstracted here to give Example \ref{ex2}. Even for a single chiral fermion there is the same shift of the KO dimension when formulating the Euclidean field theory. Physically, the parameter $i\mu$ is a Majorana mass for the left-handed fermion $\chi_L$, and $\phi_R$ is its conjugate field \cite{colemanAspectsSymmetrySelected1985}. This field is the Euclidean analogue of the Dirac conjugate of a chiral fermion in Lorentzian field theory \cite{barrettLorentzianVersionNoncommutative2007}. It is noteworthy that the one KO dimension for which the chiral integral is well defined is exactly the one needed to formulate four-dimensional physics. 
  
 \bibliographystyle{utphysmisc}

\providecommand{\href}[2]{#2}\begingroup\raggedright\endgroup
. 

\end{document}